\documentclass[conference]{IEEEtran}
\IEEEoverridecommandlockouts
\usepackage{cite}
\usepackage{amsmath,amssymb,amsfonts}
\usepackage{algorithmic}
\usepackage{graphicx}
\usepackage{textcomp}
\usepackage[table]{xcolor}
\usepackage{url}

\usepackage{breakurl}
\usepackage[breaklinks]{hyperref}
\usepackage{verbatim}
\def\BibTeX{{\rm B\kern-.05em{\sc i\kern-.025em b}\kern-.08em
    T\kern-.1667em\lower.7ex\hbox{E}\kern-.125emX}}

\newcommand{\email}[1]{\href{mailto:#1}{\texttt{#1}}}

\begin{document}

\title{A County-level Dataset for Informing the United States' Response to COVID-19}

\makeatletter
\newcommand{\linebreakand}{%
  \end{@IEEEauthorhalign}
  \hfill\mbox{}\par
  \mbox{}\hfill\begin{@IEEEauthorhalign}
}
\makeatother

\author{
\IEEEauthorblockN{Benjamin D.~Killeen\textsuperscript{*1,2}}
\IEEEauthorblockA{\email{killeen@jhu.edu}}
\and
\IEEEauthorblockN{Jie Ying Wu\textsuperscript{*1}}
\IEEEauthorblockA{\email{jieying@jhu.edu}}
\and
\IEEEauthorblockN{Kinjal Shah\textsuperscript{2}}
\IEEEauthorblockA{\email{kshah31@jhu.edu}}
\and
\IEEEauthorblockN{Anna Zapaishchykova\textsuperscript{2}}
\IEEEauthorblockA{\email{azapais1@jhu.edu}}
\linebreakand
\IEEEauthorblockN{Philipp Nikutta\textsuperscript{2}}
\IEEEauthorblockA{\email{pnikutt1@jhu.edu}}
\and
\IEEEauthorblockN{Aniruddha Tamhane \textsuperscript{1}}
\IEEEauthorblockA{\email{atamhan3@jhu.edu}}
\and
\IEEEauthorblockN{Shreya Chakraborty \textsuperscript{1}}
\IEEEauthorblockA{\email{schakr20@jhu.edu}}
\and
\IEEEauthorblockN{Jinchi Wei\textsuperscript{2}}
\IEEEauthorblockA{\email{jwei9@jhu.edu}}
\linebreakand
\IEEEauthorblockN{Tiger Gao\textsuperscript{1}}
\IEEEauthorblockA{\email{tgao11@jhu.edu}}
\and
\IEEEauthorblockN{Mareike Thies\textsuperscript{2}}
\IEEEauthorblockA{\email{mthies1@jhu.edu}}
\and
\IEEEauthorblockN{Mathias Unberath\textsuperscript{1,2}}
\IEEEauthorblockA{\email{unberath@jhu.edu}}
\linebreakand
\IEEEauthorblockA{\textsuperscript{1}\textit{Department of Computer Science} \hspace{3em}\textsuperscript{2}\textit{Laboratory for Computational Sensing and Robotics} \\
\textit{Johns Hopkins University}, \\
Baltimore, MD, United States}
\thanks{\textsuperscript{*}Equal contribution.}
\thanks{Please direct inquiries to \href{mailto:killeen@jhu.edu}{Benjamin D.~Killeen}.}
\thanks{An earlier version of this article appeared at \url{https://link.medium.com/N2azyHrq94}.}
}

\maketitle

\begin{abstract}
As the coronavirus disease 2019 (COVID-19) continues to be a global pandemic, policy makers have enacted and reversed non-pharmaceutical interventions with various levels of restrictions to limit its spread. Data driven approaches that analyze temporal characteristics of the pandemic and its dependence on regional conditions might supply information to support the implementation of mitigation and suppression strategies. 
To facilitate research in this direction on the example of the United States, we present a machine-readable dataset that aggregates relevant data from governmental, journalistic, and academic sources on the U.S. county level.
In addition to county-level time-series data from the JHU CSSE COVID-19 Dashboard~\cite{dongInteractiveWebbasedDashboard2020a}, our dataset contains more than 300 variables that summarize population estimates, demographics, ethnicity, housing, education, employment and income, climate, transit scores, and healthcare system-related metrics.
Furthermore, we present aggregated out-of-home activity information for various points of interest for each county, including grocery stores and hospitals, summarizing data from SafeGraph~\cite{safeGraph} and Google mobility reports \cite{aktay2020google}. 
We compile information from IHME, state and county-level government, and newspapers for dates of the enactment and reversal of non-pharmaceutical interventions. \cite{carterItNCBars, carroll62CoronavirusCases2020, hansenAlabamaGovernorCloses, rettnerArkansasLatestUpdates, mookBurgumClosesBars, andersonCoronavirusFloridaGovernor, ketvstaffCoronavirusNebraskaIowa2020, sogaCoronavirusVermontGovernor, kiteCoronavirusShutsKansas, feuerCoronavirusNYNJ2020, hiattAddsStayatHomeOrder2020, svitekGovGregAbbott2020, kcciGovReynoldsIssues2020, capuanoMissouriNoDiningin2020, tobinKentuckyDerbyPostponed, hnnLISTHereHow, LIVEUPDATESHere, helminiakLocalBarsResaurants2020, wgmeMaineBarsRestaurants2020, MapCoronavirusSchool2020a, ganucheauMayorsScrambleKnow2020, associatedpressMontanaExtendsSchool, etehadNevadaOrdersAll2020, whdhNewHampshireBans, krqemediaNewRestrictionsNew2020, webbPhoenixTucsonOrder2020, kludtRestaurantsBarsShuttered2020, nunesRIRestaurantsClosed2020, mervoshSeeWhichStates2020, ruskinStateBansRestaurant2020, grossStateRestrictBars, axiosStatesOrderBars, kelmanTennesseeGovernorOrders, leeTheseStatesHave, semeradUtahOrdersRestaurants, spiegelVirginiaRestaurantsBars2020, widaWhichStatesHave, star-tribuneWyomingCancellationsClosures, speciaWhatYouNeed2020, xu2020epidemiological, noauthor_coronavirus_nodate, morlan_restaurants_nodate, louie_monterey_2020, noauthor_home_nodate, noauthor_coronavirus_nodate-1, angeles_county_2020, noauthor_reopening_nodate, noauthor_san_nodate, amanda_del_castillo_revised_2020, yeager_update_nodate, noauthor_phase_nodate, noauthor_phase_nodate-1, noauthor_ron_nodate, noauthor_plan_nodate, noauthor_georgia_nodate, idaho_stages_nodate, noauthor_illinois_nodate, noauthor_update_nodate, noauthor_iowa_nodate, noauthor_gov_nodate, noauthor_updated_2020, blair_road_nodate, noauthor_whats_nodate, noauthor_baltimore_nodate, noauthor_covid-19_nodate, noauthor_covid-19_nodate-1, noauthor_coronavirus_nodate-2, noauthor_news_nodate, noauthor_prince_nodate, noauthor_safety_nodate, noauthor_minnesotas_nodate, noauthor_show_nodate, noauthor_see_nodate, noauthor_governors_nodate, noauthor_details_2020, russell_iowa_nodate, noauthor_governor_nodate, wolf24PennsylvaniaCounties2020, doxseyAnotherCOVIDCluster, COVID19InformacionGeneral, COVID19NEWSPolk, tierneyCOVID19UpdateReopening2020, COVID19UpdatesClackamas2020, raymondEverythingWeDon2020, GovHenryMcMaster2020, GovHenryMcMaster2020a, GovWolf122020, tierneyGovernorAnnouncesOrders2020, GovernorCuomoAnnounces2020, GovernorCuomoAnnounces2020a, GovernorCuomoAnnounces2020b, HundredsRestaurantsReopen, dormanKnoxCountyWill, murphyNCGettingReady, NewMexicoAllow, NorthDakotaCafes, NYCRestaurantReopening, OregonCoronavirusInformation, misincoPennsylvaniaReopeningCounties, twitterPhaseStartsTuesday, ReopeningMarionCounty, cooperRoadmapReopeningNashville, insleeSafeStartWashington2020, SaferHomePhase, richardShelbyCountyBegin, brownStateOregonNewsroom, kassahunSullivanCountyHealth2020, illersTennesseeReleasesNew, cowleyThese31Oregon, herbertUtahLeadsTogether2020, WestVirginiaStrong, WesternNewYork, WhatAllowedCounties2020, skrumWisconsinCountyList2020, WolfAnnouncesNext2020, WyomingCountyCommissioners}
By collecting these data, as well as providing tools to read them, we hope to accelerate research that investigates how the disease spreads and why spread may be different across regions. 

Our dataset and associated code are available at \href{https://github.com/JieYingWu/COVID-19_US_County-level_Summaries}{github.com/JieYingWu/COVID-19\_US\_County-level\_Summaries}.
\end{abstract}

\begin{figure}
  \centering
  \includegraphics[width=\columnwidth]{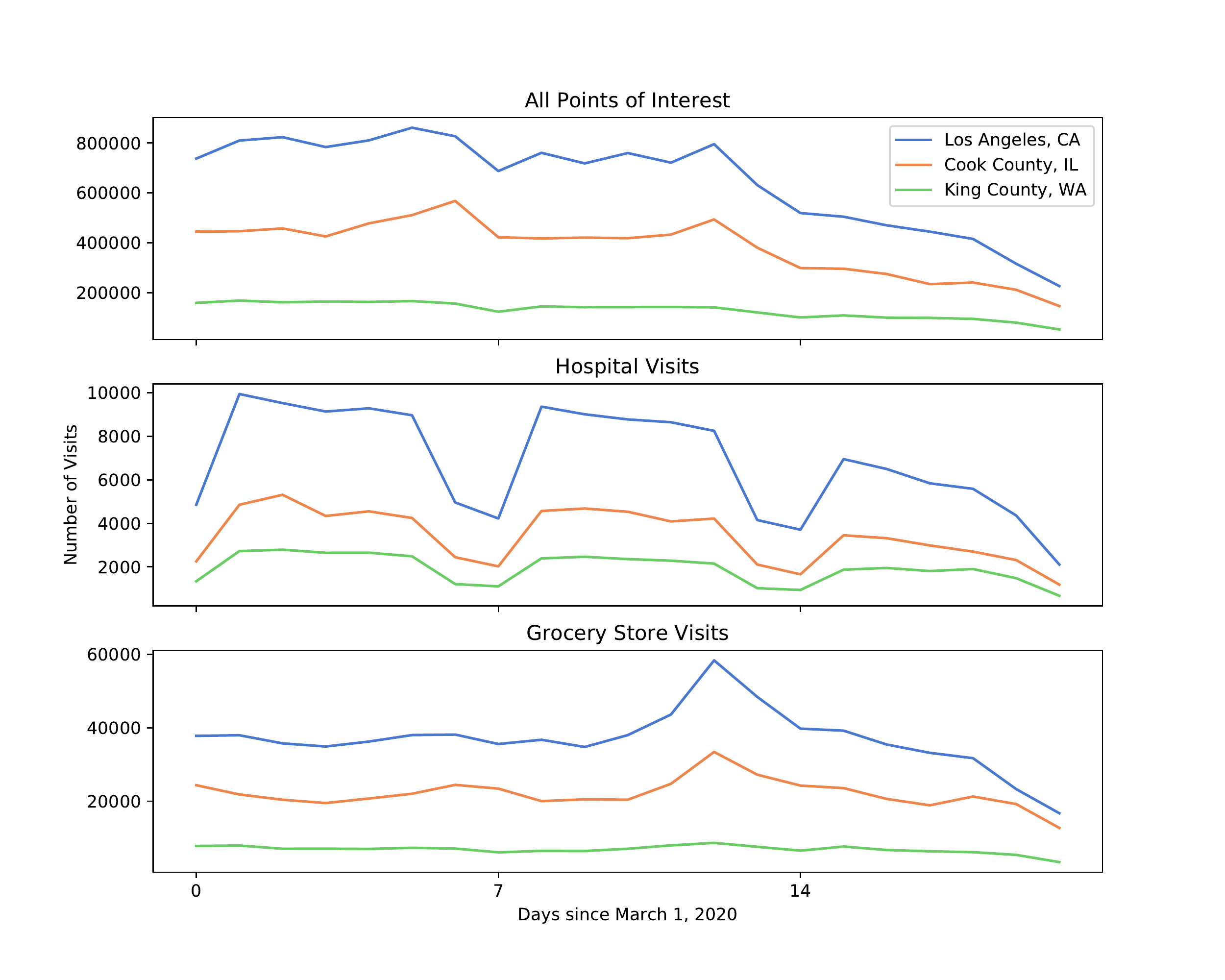}
  \caption{Aggregated out-of-home activity consisting of visits to points of interest for selected
    counties with high incidence of COVID-19 cases, from March 1 to March 21. The periodic dip in hospital visits corresponds to weekends, when most hospitals have reduced hours. The decline in overall foot traffic can be seen to start on March 12 in these counties. Data from \cite{safeGraph}.}
  \label{fig:foot-traffic}
\end{figure}

\begin{figure*}
    \centering
    \includegraphics[width=\linewidth]{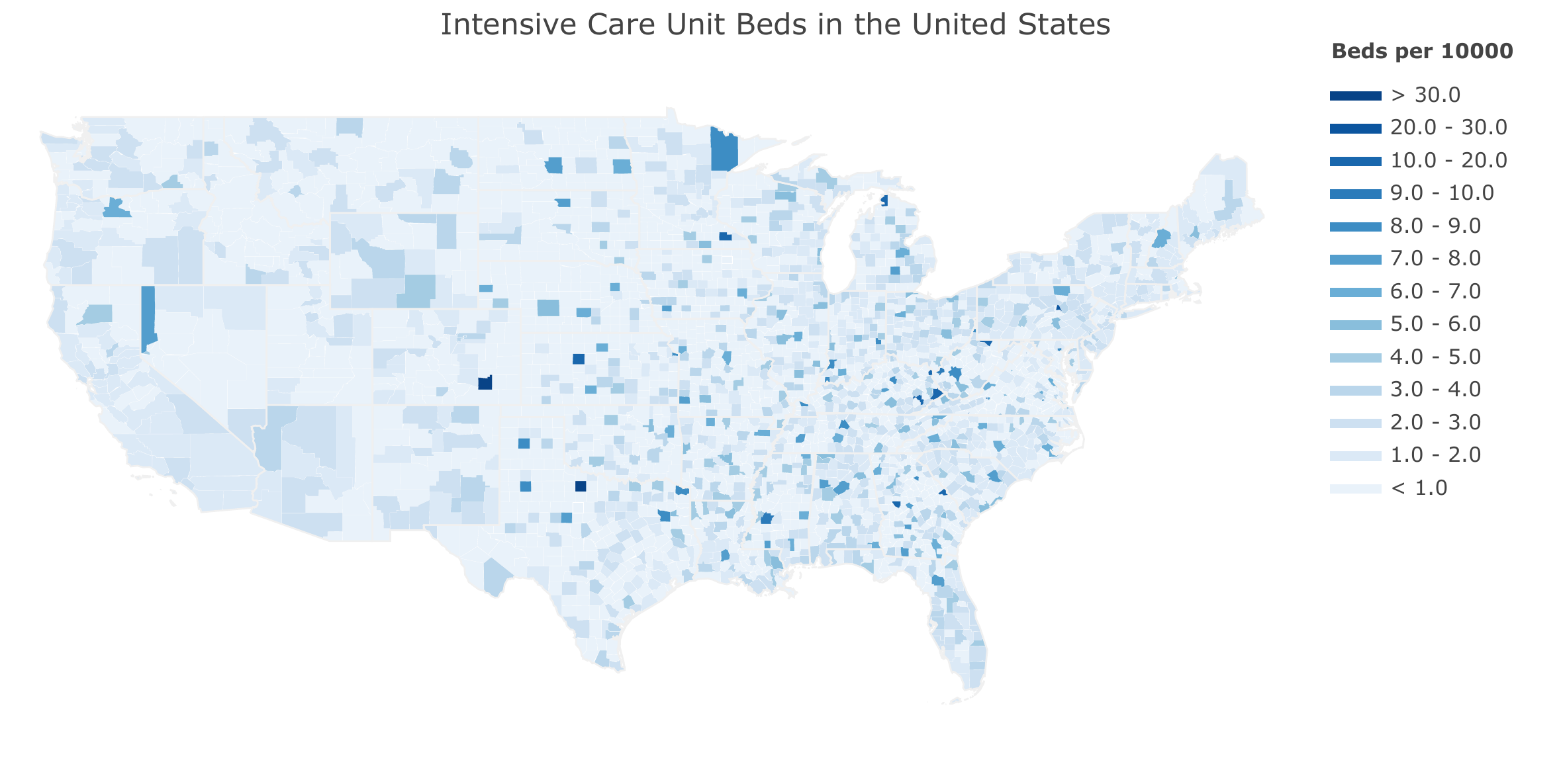}
    \vspace{-3em}
    \caption{Number of Intensive Care Unit (ICU) beds per county. Data from \cite{ICUSource}.}
    \label{fig:icu-beds}
\end{figure*}

\section{Introduction}

The ongoing outbreak of COVID-19 has had a devastating impact on the United States' health care
system, economy, and social wellbeing.  Despite early promises of an "American Resurrection" by
April 12, 2020 \cite{cnnTrumpSaysHe}, as of \today, the U.S. has unfortunately experienced more than 190,000 deaths due to COVID-19 and remains a significant epicenter of the disease with
more than 25,000 daily cases \cite{dongInteractiveWebbasedDashboard2020a, NewCasesCOVID19}. 
Social distancing measures remain in effect throughout much of the country, and despite optimistic plans to reopen schools, millions of students return to virtual classrooms this Fall due to COVID-19
\cite{hobbsMillionsStudentsHead2020}.
In many parts of the country, confirmed COVID-19 cases, hospitalizations, and deaths are increasing exponentially \cite{dongInteractiveWebbasedDashboard2020a}.
Drastic interventions like social distancing and mask mandates are necessary to slow the spread of the disease, giving more time to
\begin{itemize}
    \item provide treatment within our healthcare system's capacity,
    \item develop effective testing capability,
    \item establish sophisticated tracing mechanisms, and
    \item discover novel treatments for the virus.
\end{itemize}
At the same time, the current mitigation strategies have had severe effects on society and the economy. 
Widespread closures of schools and daycares have left working parents with limited childcare options \cite{MapCoronavirusSchool2020}; shuttered bars, restaurants, and entertainment venues have forced owners to lay off employees, predominantly in the service industry \cite{casselmanLayoffsAreJust2020}; and the U.S. and global economy may be experiencing the worst recession since World War II \cite{EmploymentSituationSummary, COVID19PlungeGlobal}.
To combat these effects, U.S. representatives have passed the largest economic stimulus package in U.S.~history \cite{HouseGivesFinal}, and the Federal Reserve has cut interest rates to near zero \cite{wesselWhatFedDoing2020}. 
However, no economic stimulus can offset the effects of altered consumer behavior.
Determining when and how to roll back non-pharmaceutical interventions in a manner which is safe and responsible is of the utmost importance.

The initial lockdown period was necessary to avoid overwhelming our hospital systems, but the current situation calls for a more nuanced approach. 
Moving forward, the U.S. must balance reducing the risk of spread with the adverse economic consequences of millions of furloughed and unemployed people. To inform this process, we have curated a \textbf{machine-readable dataset} that aggregates data from governmental, journalistic, and academic sources on the county level, including aggregated NPI implementation dates. While most of these sources are freely available, there is substantial work to align them and put them in a standard format that enables analysis.
In addition to time-series data from \cite{dongInteractiveWebbasedDashboard2020a}, which details COVID-19 per-county infections and deaths, our dataset contains more than 300 variables that summarize population estimates, demographics, ethnicity, housing, education, employment and income, climate, transit scores, and healthcare system-related metrics.
Further, we source a significant number of journal articles and official statements detailing implementation dates of interventions, including mask mandates, stay-at-home orders, school closures, and restaurant and entertainment venue closures \cite{carterItNCBars, carroll62CoronavirusCases2020, hansenAlabamaGovernorCloses, rettnerArkansasLatestUpdates, mookBurgumClosesBars, andersonCoronavirusFloridaGovernor, ketvstaffCoronavirusNebraskaIowa2020, sogaCoronavirusVermontGovernor, kiteCoronavirusShutsKansas, feuerCoronavirusNYNJ2020, hiattAddsStayatHomeOrder2020, svitekGovGregAbbott2020, kcciGovReynoldsIssues2020, capuanoMissouriNoDiningin2020, tobinKentuckyDerbyPostponed, hnnLISTHereHow, LIVEUPDATESHere, helminiakLocalBarsResaurants2020, wgmeMaineBarsRestaurants2020, MapCoronavirusSchool2020a, ganucheauMayorsScrambleKnow2020, associatedpressMontanaExtendsSchool, etehadNevadaOrdersAll2020, whdhNewHampshireBans, krqemediaNewRestrictionsNew2020, webbPhoenixTucsonOrder2020, kludtRestaurantsBarsShuttered2020, nunesRIRestaurantsClosed2020, mervoshSeeWhichStates2020, ruskinStateBansRestaurant2020, grossStateRestrictBars, axiosStatesOrderBars, kelmanTennesseeGovernorOrders, leeTheseStatesHave, semeradUtahOrdersRestaurants, spiegelVirginiaRestaurantsBars2020, widaWhichStatesHave, star-tribuneWyomingCancellationsClosures, speciaWhatYouNeed2020}, as well as reopenings \cite{xu2020epidemiological, noauthor_coronavirus_nodate, morlan_restaurants_nodate, louie_monterey_2020, noauthor_home_nodate, noauthor_coronavirus_nodate-1, angeles_county_2020, noauthor_reopening_nodate, noauthor_san_nodate, amanda_del_castillo_revised_2020, yeager_update_nodate, noauthor_phase_nodate, noauthor_phase_nodate-1, noauthor_ron_nodate, noauthor_plan_nodate, noauthor_georgia_nodate, idaho_stages_nodate, noauthor_illinois_nodate, noauthor_update_nodate, noauthor_iowa_nodate, noauthor_gov_nodate, noauthor_updated_2020, blair_road_nodate, noauthor_whats_nodate, noauthor_baltimore_nodate, noauthor_covid-19_nodate, noauthor_covid-19_nodate-1, noauthor_coronavirus_nodate-2, noauthor_news_nodate, noauthor_prince_nodate, noauthor_safety_nodate, noauthor_minnesotas_nodate, noauthor_show_nodate, noauthor_see_nodate, noauthor_governors_nodate, noauthor_details_2020, russell_iowa_nodate, noauthor_governor_nodate, wolf24PennsylvaniaCounties2020, doxseyAnotherCOVIDCluster, COVID19InformacionGeneral, COVID19NEWSPolk, tierneyCOVID19UpdateReopening2020, COVID19UpdatesClackamas2020, raymondEverythingWeDon2020, GovHenryMcMaster2020, GovHenryMcMaster2020a, GovWolf122020, tierneyGovernorAnnouncesOrders2020, GovernorCuomoAnnounces2020, GovernorCuomoAnnounces2020a, GovernorCuomoAnnounces2020b, HundredsRestaurantsReopen, dormanKnoxCountyWill, murphyNCGettingReady, NewMexicoAllow, NorthDakotaCafes, NYCRestaurantReopening, OregonCoronavirusInformation, misincoPennsylvaniaReopeningCounties, twitterPhaseStartsTuesday, ReopeningMarionCounty, cooperRoadmapReopeningNashville, insleeSafeStartWashington2020, SaferHomePhase, richardShelbyCountyBegin, brownStateOregonNewsroom, kassahunSullivanCountyHealth2020, illersTennesseeReleasesNew, cowleyThese31Oregon, herbertUtahLeadsTogether2020, WestVirginiaStrong, WesternNewYork, WhatAllowedCounties2020, skrumWisconsinCountyList2020, WolfAnnouncesNext2020, WyomingCountyCommissioners} .
Finally, we aggregate out-of-home activity data from \cite{safeGraph} and \cite{aktay2020google} in each county, possibly measuring compliance with the aforementioned restrictions.
Fig.~\ref{fig:foot-traffic} shows a sample of out-of-home activity for selected counties.

We hope that this dataset proves to be a useful resource to the community, facilitating important research on epidemiological forecasting.
In particular, a machine learning approach to identify highly relevant factors may inform a graduated rollback of isolation measures and travel restrictions.

\begin{table*}[ht]
    \centering
    \rowcolors{2}{gray!25}{white}
    \begin{tabular}{|l|l|l|}
        \hline
        \rowcolor{gray!50}
        Data Type & Source & Availability \\
        COVID-19 Infections COVID-19 Related Deaths Time-series & \cite{dongInteractiveWebbasedDashboard2020a} & ---\\
         2020 Date of COVID-19 Interventions, \textit{e.g.}~stay-at-home order & 
        \cite{carterItNCBars, carroll62CoronavirusCases2020, hansenAlabamaGovernorCloses, rettnerArkansasLatestUpdates, mookBurgumClosesBars, andersonCoronavirusFloridaGovernor, ketvstaffCoronavirusNebraskaIowa2020, sogaCoronavirusVermontGovernor, kiteCoronavirusShutsKansas, feuerCoronavirusNYNJ2020, hiattAddsStayatHomeOrder2020, svitekGovGregAbbott2020, kcciGovReynoldsIssues2020, capuanoMissouriNoDiningin2020, tobinKentuckyDerbyPostponed, hnnLISTHereHow, LIVEUPDATESHere, helminiakLocalBarsResaurants2020, wgmeMaineBarsRestaurants2020, MapCoronavirusSchool2020a, ganucheauMayorsScrambleKnow2020, associatedpressMontanaExtendsSchool, etehadNevadaOrdersAll2020, whdhNewHampshireBans, krqemediaNewRestrictionsNew2020, webbPhoenixTucsonOrder2020, kludtRestaurantsBarsShuttered2020, nunesRIRestaurantsClosed2020, mervoshSeeWhichStates2020, ruskinStateBansRestaurant2020, grossStateRestrictBars, axiosStatesOrderBars, kelmanTennesseeGovernorOrders, leeTheseStatesHave, semeradUtahOrdersRestaurants, spiegelVirginiaRestaurantsBars2020, widaWhichStatesHave, star-tribuneWyomingCancellationsClosures, speciaWhatYouNeed2020} &
        --- \\
        2020 Date of COVID-19 Intervention Rollbacks & \cite{xu2020epidemiological, noauthor_coronavirus_nodate, morlan_restaurants_nodate, louie_monterey_2020, noauthor_home_nodate, noauthor_coronavirus_nodate-1, angeles_county_2020, noauthor_reopening_nodate, noauthor_san_nodate, amanda_del_castillo_revised_2020, yeager_update_nodate, noauthor_phase_nodate, noauthor_phase_nodate-1, noauthor_ron_nodate, noauthor_plan_nodate, noauthor_georgia_nodate, idaho_stages_nodate, noauthor_illinois_nodate, noauthor_update_nodate, noauthor_iowa_nodate, noauthor_gov_nodate, noauthor_updated_2020, blair_road_nodate, noauthor_whats_nodate, noauthor_baltimore_nodate, noauthor_covid-19_nodate, noauthor_covid-19_nodate-1, noauthor_coronavirus_nodate-2, noauthor_news_nodate, noauthor_prince_nodate, noauthor_safety_nodate, noauthor_minnesotas_nodate, noauthor_show_nodate, noauthor_see_nodate, noauthor_governors_nodate, noauthor_details_2020, russell_iowa_nodate, noauthor_governor_nodate, wolf24PennsylvaniaCounties2020, doxseyAnotherCOVIDCluster, COVID19InformacionGeneral, COVID19NEWSPolk, tierneyCOVID19UpdateReopening2020, COVID19UpdatesClackamas2020, raymondEverythingWeDon2020, GovHenryMcMaster2020, GovHenryMcMaster2020a, GovWolf122020, tierneyGovernorAnnouncesOrders2020, GovernorCuomoAnnounces2020, GovernorCuomoAnnounces2020a, GovernorCuomoAnnounces2020b, HundredsRestaurantsReopen, dormanKnoxCountyWill, murphyNCGettingReady, NewMexicoAllow, NorthDakotaCafes, NYCRestaurantReopening, OregonCoronavirusInformation, misincoPennsylvaniaReopeningCounties, twitterPhaseStartsTuesday, ReopeningMarionCounty, cooperRoadmapReopeningNashville, insleeSafeStartWashington2020, SaferHomePhase, richardShelbyCountyBegin, brownStateOregonNewsroom, kassahunSullivanCountyHealth2020, illersTennesseeReleasesNew, cowleyThese31Oregon, herbertUtahLeadsTogether2020, WestVirginiaStrong, WesternNewYork, WhatAllowedCounties2020, skrumWisconsinCountyList2020, WolfAnnouncesNext2020, WyomingCountyCommissioners} & --- \\
        March, 2020 Out-of-home Activity Time-series & 
        SafeGraph \cite{safeGraph} & --- \\
        2018 Population Estimates &
        Census \cite{PopEst} & 97-100\% \\
        2014-2018 Educational Attainment &
        Census \cite{EduAttain} & 100\% \\
        2018 Estimated Poverty Level & USDA~\cite{EstPoverty} & 97\%\\
        2018 Employment and Income & USDA~\cite{EmpInc} & 99\% \\
        2019 Precipitation and Temperature &
        NOAA~\cite{PrecTemp} & 86\% (37.8\% imputed)\\
        2010 Housing and Density & 
        Census \cite{HouseLand} & 99\% \\
        2018 Age Group Demographics &
        Census \cite{HouseDemo} & 97\%\\
        2018 Household Demographics &
        Census \cite{HouseDemo} & 25\%\\
        2018 Ethnic Group Demographics &
        Census \cite{Ethnic} & 97\%\\
        2019 Healthcare Capacity: Physicians, NPs, PAs & 
        AAMC, KFF \cite{AAMCPhy, PrimaryCareDoctorsSource, SpecialistDoctorsSource} & 86-97\%\\
        2019 Healthcare Capacity: ICU Beds  & KFF \cite{hospitals, ICUSource} & 92-97\%\\
        2019 Public Transit Scores & 
        CNT \cite{PubTrans} & 95\%\\
        2016 Crime Rates & 
        DOJ \cite{CrimeStats} & 97\%\\
        \hline
    \end{tabular}
    \vspace{1em}
    \caption{Data Source Descriptions and Percentage of Counties Included For Static Data}
    \label{tab:data-summary}
\end{table*}

\section{Related Work}

Because of the rapidly-evolving nature of the COVID-19 pandemic, the response from the data science community is ongoing and in flux. Here, we review related efforts which were influential at the outset of the pandemic. As new articles are published every day, this is by no means an exhaustive review.

Despite significant public interest, government agencies have yet to publish a county-level data source for cases of COVID-19. The World Health Organization has gathered self-reported data on the national level \cite{WHOCoronavirusDisease}, while the United States Center for Disease Control reports state-level infection and fatality rates \cite{cdcCoronavirusDisease20192020}. However, \cite{dongInteractiveWebbasedDashboard2020a, timesWeReSharing2020} continue to maintain the most up-to-date and reputable collections of COVID-19 cases across the United States, hosted by the Center for Systems Science and Engineering (CSSE) at Johns Hopkins University and the New York Times, respectively.
These efforts focus on current, hard data gathered from local government publications and reputable journalistic sources.
Other efforts focus on gleaning related information from a variety of sources, including social media. 
\cite{chenCOVID19FirstPublic2020} tracks COVID-19 related tweets in an effort to understand the conversation and possible misinformation surrounding the pandemic. Johns Hopkins University, University of Maryland, and George Washington University have also started a collaboration to track COVID-19 through social media \cite{socialMedaForPublicHealth}.

A large body of work has focused on using machine learning and data science tools to understand the virus.
\cite{zhangEstimationReproductiveNumber2020} uses data from the Diamond Princess cruise ship, where an early outbreak took place, to estimate the reproductive number $R_0$ of the virus.
\cite{santoshAIDrivenToolsCoronavirus2020} implements active learning methods to detect new outbreaks of the virus, incorporating new data types without having to retrain.
\cite{fongCompositeMonteCarlo2020, fongFindingAccurateEarly2020} focus on understanding the current pandemic in its early stages, compensating for the inherent uncertainty in novel disease.
Finally, \cite{flaxman2020estimating} applies a data-driven approach to understand the effect of NPIs on the reproductive ratio of COVID-19 in European countries.

\section{Dataset}

We describe the structure of our dataset, which includes each component in its raw form as well as a narrowed-down, machine-readable form conducive to a machine-learning approach. 
Table~\ref{tab:data-summary} summarizes the sources and availability for each type of data, and a full description of each variable can be found in our repository.

\subsection{County Descriptors}
\label{sec:county-descriptors}

We populate a CSV file with over 300 variables for 3220 county-equivalent areas (as well as the fifty states, District of Columbia, and the whole United States) with numerous types of data, including population, education, economic, climate, housing, health care capacity, public transit, and crime statistics.
Each area is uniquely identified by its Federal Information Processing Standard (FIPS) code, a five digit number where they first two digits designate the state, and the last three digits describe the county-equivalent.
Our sources include the United States Census Bureau \cite{PopEst, EduAttain, HouseDemo, Ethnic}, the United States Department of Agriculture (USDA) Economic Research Service \cite{EstPoverty, EmpInc}, the National Oceanic and Atmosphere Administration (NOAA) \cite{PrecTemp}, the Association of American Medical Colleges (AAMC) \cite{AAMCPhy}, the Henry J.~Kaiser Family Foundation (KFF) \cite{PrimaryCareDoctorsSource, SpecialistDoctorsSource, ICUSource}, the Center for Neighborhood Technology (CNT) \cite{PubTrans}, and the Bureau of Justice Statistics, Department of Justice (DOJ) \cite{CrimeStats}.
Perhaps most relevant to the ongoing effort to mitigate the effects of COVID-19 in the U.S.~is county-level healthcare system capacity.
The dataset includes detailed counts for each type of medical practitioner as well as the number of Intensive Care Unit beds in each county, shown in Fig.~\ref{fig:icu-beds}.

For the most part, these basic descriptive variables are unaltered from their original state. Where appropriate, missing values have been imputed with the state-wide average, detailed in Table~\ref{tab:data-summary}.


\begin{figure*}
    \centering
    \includegraphics[width=0.99\linewidth]{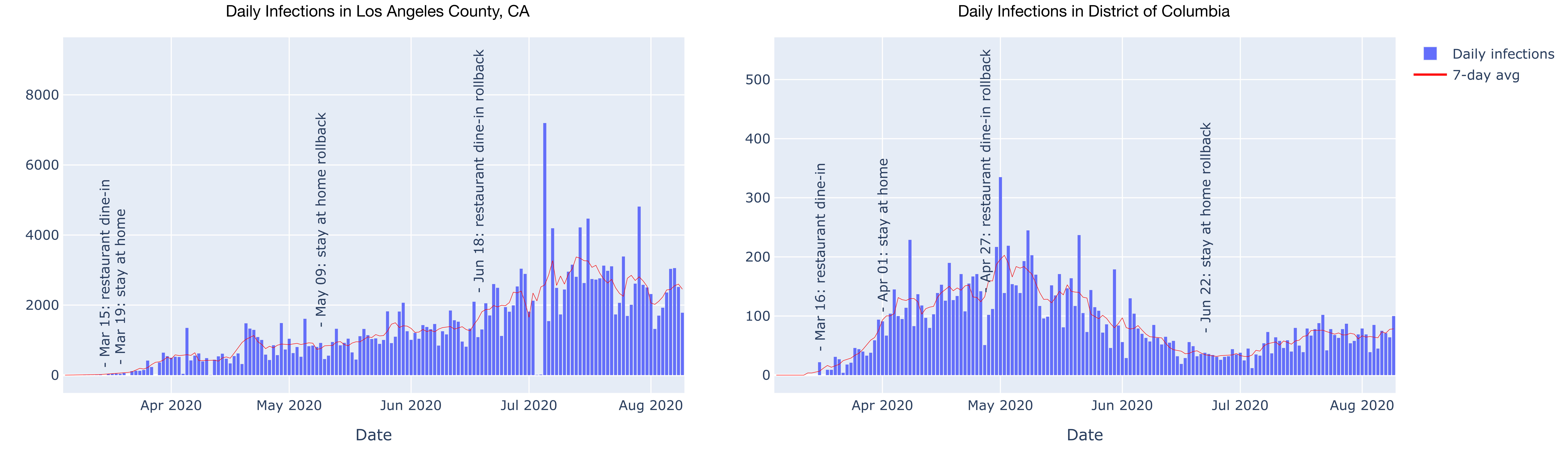}
    \caption{The daily number of infections for Los Angeles County and the District of Columbia, \cite{dongInteractiveWebbasedDashboard2020a}. The 7-day average is shown in red. For simplicity, we only show the implementation dates for two interventions and their rollbacks, stay-at-home orders and restaurant closures, since many interventions overlap. Note that the scales differ by a factor of 20.}
    \label{fig:infections-la-dc}
\end{figure*}

\subsection{Interventions}
Our dataset describes mitigation efforts taken at the state level, including stay-at-home advisories, banning large gatherings, public school closures, and restaurant and entertainment venue closures. We also include the rollback of these mitigation efforts up to Aug 2nd. 
For machine readability, we provide each date of implementation as a Gregorian ordinal, \textit{i.e.} the integer number of days starting at January 1, Year 1 CE, consistent with standard software libraries.
Moreover, these data are provided according to the same county-level row ordering as our county descriptor data (see Sec.~\ref{sec:county-descriptors}). 
Interventions made at the state level have been assigned to each county in that state, and we include county-level interventions wherever possible.
An intervention is designated \texttt{NA} if the county or state has not yet enacted it. For the most parts, county-level NPI implementation was gathered through local newspapers and government websites. The full list of these sources can be found \url{https://github.com/JieYingWu/COVID-19_US_County-level_Summaries/tree/master/data}. Since rollbacks were implemented in a more staggered fashion, we started by using the dates provided by the IHME database, which contains state-level rollback information.~\cite{xu2020epidemiological} While IHME records the categories for stay-at-home and gatherings, they separate businesses as essential and non-essential. We use dates for non-essential businesses reopening for restaurants and gym/entertainment. As schools were mostly still on summer vacation by Aug 2nd, we did not collect school reopenings. To refine the state-level data to the county level, we use reopening information from the New York Times\cite{leeSeeHowAll2020}, which includes some counties that do not follow state guidelines. We also rely on it to fix discrepancies such as if the opening of the first non-essential business did not include restaurants or gyms. Additionally, since counties are generally driven to implement a different rollback schedule if they have an unusual number of COVID-19 cases, we check the county government website of those counties where there was a drastic uptick in cases in the previous months. As different counties have reopened at different levels, such as reopening restaurants at 25\% outdoor seating, we count any amount of reopening as that NPI has been rolled back. As the policies surrounding COVID-19 management is continually changing, we appreciate any contributions to the repository to keep it up-to-date, especially as school reopening decisions come into effect. 



\subsection{Out-of-home Activity and Mobility}

We have aggregated point-of-interest location data gathered from user's smartphones to show out-of-home activity, using raw data from \cite{safeGraph}.
For privacy and IP reasons, our dataset does not include user location data in its raw form but rather in several time-series files summarizing county-level activity.
Fig.~\ref{fig:foot-traffic} shows the time-series for selected counties which have a high incidence of COVID-19 cases.
The decline in overall activity on March 12 corresponds to an increased media attention and stay-at-home advisories in those areas.
At the same time, a spike in grocery store visits points to a panic-buying spree which has since subsided.

Additionally we include data from Google mobility reports \cite{aktay2020google}, which may correlate with changes in the reproductive ratio of COVID-19.
These include aggregated and anonymized data, detailing the percent change in number of visits to six location types compared to baseline: grocery and pharmacy; parks; residential; retail and recreation; transit stations; and workplaces.
Visits to residential addresses likely describes individuals staying at home, with the obvious exception of gatherings that occur at residential addresses, either for work or social reasons.

\subsection{Disease Spread}
Finally, we provide time-series data for the cumulative number of COVID-19 confirmed cases and related deaths, from \cite{dongInteractiveWebbasedDashboard2020a}.
This data begins on January 22, 2020.
It should be noted that epidemiological modeling efforts may want to consider the uncertainty surrounding U.S. testing \cite{scherUSSeverelyUndertesting}, on which these data are based.
At the time of this writing, efforts to improve the availability of COVID-19 tests are ongoing, but the current strategies prioritize patients with severe symptoms.
Thus, modeling efforts may wish to take into account random subsampling of the true population, where untested individuals still spread the virus.
This is especially true given that nearly half of all COVID-19 infections may be asymptomatic \cite{CoronavirusCasesSymptoms2020}. 


Fig.~\ref{fig:infections-la-dc} shows the daily number of infections in Los Angeles County, CA, and the District of Columbia according to \cite{dongInteractiveWebbasedDashboard2020a}, as well as implementation dates for select interventions: ``stay at home'' and ``restaurant dine-in.''
In some areas, rollbacks have coincided with a resurgence of the virus, reaching levels of new daily infections far greater than the initial outbreak, as in Los Angeles, whereas other areas have rolled back NPIs and experienced only a small or negligible increase, such as the District of Columbia (see Fig.~\ref{fig:infections-la-dc}).

%

\section{Discussion}

The resurgence of COVID-19 in some areas but not others reinforces the need for continued vigilance everywhere.
In some sense, the United States has experienced not a single outbreak but multiple outbreaks, both simultaneous and non-simultaneous, with differing characteristics in terms of transmission rate, mobility, and response to NPIs.
Although pharmaceutical interventions, such as a vaccine, or natural herd immunity may eventually mitigate the likelihood of an outbreak independent from public behavior, these eventualities are still far on the horizon.
In the meantime, the possibility of a resurgence, which may overwhelm the healthcare systems, is ever-present. 
The number of individuals who will ultimately be infected---and the number of deaths that will result---depend on the interventions reinforced now.
At the same time, the economic impact of these interventions, which is not evenly distributed across counties, cannot be ignored.
It depends on the characteristic qualities of each area---very different, for example, New York as opposed to Silicon Valley.
The former has a large population in the entertainment and service industries, which will need financial support during quarantine, whereas the latter is dominated by large tech firms, whose employees can adapt to working from home.
By providing the socioeconomic attributes of each county, the spread of COVID-19 confirmed cases, and the ongoing response in a machine-readable format, we hope to inform the decisions made to most effectively protect each area.

\section*{Acknowledgment}

Thank you to all our sources, especially the JHU CSSE COVID-19 Dashboard for making their data public and SafeGraph, for providing researchers their data for COVID-19 related work.

\bibliographystyle{IEEEtran}
\bibliography{references.bib}

\begin{thebibliography}{100}
\bibitem{safeGraph}
SafeGraph, ``Footprint data,'' safeGraph, a data company that aggregates
  anonymized location data from numerous applications in order to provide
  insights about physical places. To enhance privacy, SafeGraph excludes census
  block group information if fewer than five devices visited an establishment
  in a month from a given census block group.
\end{thebibliography}

\end{document}